\documentstyle[pre,aps,multicol,epsfig,color]{revtex}
  
\begin{document}
\draft

\title{Viability of an elementary syntactic structure in a population playing Naming Games} 

\author{Edgardo Brigatti}  
  
\address{Instituto de F\'{\i}sica, Universidade Federal do Rio de Janeiro, 
Av. Athos da Silveira Ramos, 149,
Cidade Universitária, 21941-972, Rio de Janeiro, RJ, Brasil}
%\address{Instituto de F\'{\i}sica, Universidade Federal Fluminense, 
%  Campus da Praia Vermelha, 24210-340, Niter\'oi, RJ, Brazil}
\address{e-mail address: edgardo@if.ufrj.br}

\date{\today}
\maketitle
\widetext
  
\begin{abstract}

 We explore how the social dynamics of communication
and learning can bring about the rise 
of a syntactic communication in a 
population of speakers.
Our study is developed  starting
from a version of the Naming Game model
where an elementary syntactic structure is introduced.
This analysis shows how the transition from non-syntactic to syntactic 
communication is socially favored in %small 
communities which need to 
exchange a large number of concepts.

 \end{abstract}
  

\begin{multicols}{2}

\section{Introduction}

The study of the evolution of languages and their structures has generated
a very rich debate crossing different disciplines and approaches.

The ideas developed in the linguistic community, 
which introduced the thesis of considering 
some linguistic structures as innate with some specific properties genetically
encoded in a language module or organ \cite{chomsky1} have been very 
prolific. They bootstrapped the development of many works 
where pure evolutionary perspectives are introduced to
explain the generation of languages.
Here, the dominant paradigm is the Darwinian evolution of
biological systems.
The description of language evolution is based on a biological dynamics 
constructed  above the concepts of natural selection, mutation, and fitness, 
elaborated in terms of communication success \cite{nowak}.
This approach is particularly well suited to describe evolution from a 
functionalist perspective, where the category of utility is the one that drives the dynamics.
In any event,  many linguistic properties appear to be so highly abstract as to even hinder communication \cite{chomsky05}. This means that they are quite difficult to be introduced 
on purely functional basis  and they can not be explained merely in terms of communicative effectiveness or cognitive constraints.
Moreover, it is hard to explain how shifts from learned linguistic conventions can be fixed into genetically encoded principles necessary to evolve a language module.
Cultural conventions change much more rapidly than genes and the Baldwin effect, 
a possible Darwinian solution to this challenge, can not be the solution to this puzzle \cite{chater1}.
Biological models can be seen more as a powerful metaphor for studying the effects 
of random copying and selection, but more specific mechanisms, typically related to cultural transmission, should be considered.
 
Recently, attention was paid to defining specific 
cultural dynamics, directly related to linguistic ability.
The mechanisms which define the dynamics of the evolution of
language are different from those underlying biological evolution. 
Language is transmitted among people through learning and not DNA.
It is shaped by processes of cultural transmission across generations
of language learners.
In this view, linguistic constructions are not innate, 
but rather they are acquired through some form of probabilistic learning.
This learning process, articulated on the use of 
cognition-general principles, has become the central issue governing  
language evolution.
In fact, learning defines the dynamics of linguistic variants
and  the differences among language learnability  control these dynamics.
Learnability is quantified measuring
the learner's capacity to recover a complete description 
of  a linguistic construction to which she/he has been exposed sufficiently \cite{hsu}. 
Several works have studied the differential learnability of competing linguistic 
variant \cite{pearl,christiansen}, and also their dynamics in the absence 
of selection \cite{reali}.

Biological dynamics and cultural dynamics of learning are 
two central issues which determine the creation of linguistic structures,
but they are not the only ones.
Language is constructed for communication. 
It is not only the basis for social relationships,
but it is also based on social relationships. 
Individual learning is just one aspect of a
more general and collective process.
The fixation of linguistic conventions among a population of speakers 
is another dynamics related to the linguistic definition, and 
the structures which appear to be learnable at an individual level must be 
fixed socially. 
These social dynamics cause a pressure on language, 
which shapes a shared communication system. 
This is  a form of  collective learning.
It is important  to perceive how on a social level 
even completely arbitrary linguistic properties can succeed.
In fact, if the same convention is adopted by all members 
of a community
this convention can work and finally it becomes fixed.
The only important fact is
that everyone adopts the same 
set of culturally mediated conventions.
Even different conventions, if equally effective, may serve equally well
if there are no costs or no conflicting functional pressures.
Fixation of structures is not driven by a fitness or a learnability advantage, 
but rather by the mechanisms which generate consensus 
about the linguistic elements used by speakers.
This process is not necessarily a functional process and is not
only driven by utility. 
A lot of works related to this social aspect of language appeared recently, 
following a seminal paper by L. Steels \cite{steels2}.
These ideas were first used to describe the
birth of neologisms, and they have been tested by an artificial experiment in which
embodied software agents bootstrap a shared lexicon 
without any external intervention  \cite{steels1}. 
Robots concretize a {\it language game} 
developing a vocabulary throughout a self-organized process called Naming Game. 
Recently, these studies have also attracted the interest of the 
Statistical Physics community.
An initial study in this direction \cite{baronca}, 
was inspired by experiments conducted with the use of robots \cite {steels2}.
In that work each player is characterized by an inventory of words which can all
be used to name the same object.
At each time step two players, randomly chosen, interact following 
some simple rules.
These dynamics force the system to undergo a disorder/order transition 
towards an absorbing state characterized by a common word used by 
all the players. 
These ideas have been developed to describe other phenomena, such as
 the emergence of universality in color naming patterns \cite{gong} 
 and the self-organization 
 of a hierarchical category structure  in a population of individuals \cite{puglisi}.\\ 

The ideas and considerations  we have exposed can be applied 
across different linguistic levels:  lexicon, phonology, morphology, and syntax.
In this work, we will focus our attention on syntax.
Syntax is a process to combine progressively 
symbolic units in an ordered 
output which falls within the quite narrow bounds that delimit human language.
This is obtained by merging words into larger units and superimposing algorithms that determine 
the reference of items that might otherwise be ambiguous or misleading \cite{bicke1}.
As proposed in the Minimalist Program \cite{chomsky3}, the basic syntax-creating process 
is Merge, a process that takes two units (words, phrases, clauses) and forms them 
into a single one satisfying some constraints. 
This means that Merge has many restrictions on the items to be merged, and there is a consistent way of merging %(combining) 
them.
 As we are interested in the transition from non-syntactic to syntactic communication it is reasonable to look for the
simplest   advance %evolution progress 
from the pre-syntactic (one-word) stage, even if it does
not specifically correspond to the syntax of some 
present-day language.
This first step can be identified with the
most basic (proto-)syntactic combination:
flat concatenation of two symbols, 
where all the possible combinations are functional \cite{progovac}.
%This is clearly the most basic (proto-)syntactic combination.
 This correspond to a
purely linear bead-stringing process, %are operating.
%This 
a practice which underlies protolanguages, like the one used by speakers of a pidgin language \cite{bickerton}.
%These compounds involve a proto-Merge, that is, Merge that does not create hierarchical structure, and can be identified as the very foundation of syntax.

In our work we are interested in exploring the transition from 
non-syntactic to syntactic communication from a social dynamics 
point of view. 
Directly following  the ideas of Nowak et al. \cite{nowak2}, the 
example that we are going to explore can be stated in this way.
Let us consider the situation where a speaker is interested in communicating some concepts.
If she/he uses a non-syntactic language a symbol (word) is used  for each concept.
In the case of a syntactic language a combination of two symbols, for example
one for the object %(noun) 
$i$ and one for the action %(verb) 
$j$, can be used  to
communicate the concept $C_{ij}$ \cite{nowak2}.  
In the following we will consider the simplified situation 
where the number of object and action exchanged in the communication are the same ($S$).
Moreover, all the possible combinations of these symbols can occur and correspond to a meaningful concept.
It follows that, in this model, the possible combinations can be represented by a
matrix $C_{ij}=S\times S$ (see Table~\ref{tabula}).
For example, as a particular situation we can think that the lines elements represent nouns and the column elements verbs.
A population of individuals coevolve this system of symbols, with or without syntax, by 
playing elementary language games (Naming Game)  analogous to the ones introduced in \cite{baronca}.
In this way we can analyze the differences between syntactic and non syntactic 
communication %on the basis of the social communicative potential of a linguistic structure,
and we can distinguish when the transition from non-syntactic to syntactic 
communication is socially favored. 
\\

\begin{table}
\begin{center}
\begin{tabular}{c|ccc}
& a & b & c \\ 
\hline
d & a+d & b+d & c+d \\ 
e & a+e & b+e & c+e  \\ 
f & a+f & b+f & c+f \\ 
\end{tabular}
\end{center}
%%\noindent
\label{tab_Con}
\caption{ \small In a Naming Game with syntax for the communication of $9$ concepts
we represent these concepts using the matrix $C=3X3$. 
Each concept is specified by the couple formed by two different possible words 
contained in two different  inventories among the six inventories $a,b,c,d,e,f$.
For example, in a concrete situation $a,b,c$ can stand for an object (noun) and $d,e,f$ for an action (verb).
}
\label{tabula}
\end{table}

The paper is organized as follows. The Section II.A introduces
a version of the basic Naming Game model for the communication of 
one concept and Section II.B illustrates the model with one concept and
syntax. 
Section II.C describes the generalization of the model for 
many different concepts, in the case of a syntactic or non-syntactic communications. 
In Section III.A we show the numerical results obtained for a one concept model 
using syntax, or not, along the communications.
Section III.B is devoted to illustrating what happens with the introduction of syntax in a 
many-concept Game. Conclusions are reported in Section IV. 

\section{The model}
\subsection{The basic model for one concept}

The Naming Game is played by $P$ agents who try to reach consensus
in naming a single concept. An inventory, which 
contains an arbitrary number of words, represents each agent.
Population starts with empty inventories.
At each time step, two agents are randomly selected;
the first one assumes the role of speaker, the second one 
of hearer.
Then, the following microscopic rules \cite{baronca},  
control their actions:

1) The speaker retrieves a word 
from its inventory or, if its inventory is empty, invents a new word.

2) The speaker transmits the selected word to the hearer.

3a) If the hearer's inventory contains such a word, 
the communication is a success. 
The two agents update their inventories so as to keep 
only the word involved in the interaction. 
 
3b) Otherwise the communication is a failure. 
The hearer learns the word communicated by the speaker.

\begin{figure}
\begin{center}
\includegraphics[width=0.5\textwidth, angle=0]{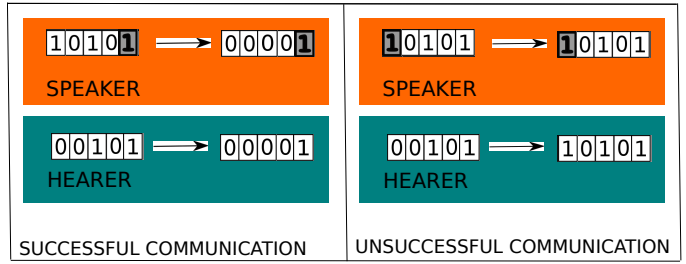}
\end{center}
\caption{\small  
(Color online)
Example of the dynamics of the inventories for a one concept model
without syntax. On the left we present a successful Game, on the right a failed one. 
We represent the 32 possible symbols with only five slots for the sake of clarity.
When a symbol is not present in the inventory the corresponding bit is set to 
0, otherwise, when the symbol is present, the bit is set to 1. 
The shadowed elements are the ones transmitted by the speaker.
}
\label{fig_draw0}
\end{figure}

The players invent new words choosing among $32$ 
possible ones with equal probability.
In contrast with the classical implementation of the model 
\cite{baronca} we work with a fixed maximum number of possible 
different words.
This is the only difference between our implementation
and the classical one. 
We introduce this simplification with the goal of
implementing a light model that can be easily generalized for the description of the 
naming process for more than one concept. 
For this reason we need a fast algorithm.
This is obtained by using boolean 
programming techniques which causes a veritable improvement  
in the computational times.
An example of these dynamics 
is represented in Figure~\ref{fig_draw0}.

\subsection{The syntax model for one concept}

Now we consider a context in which
one single concept is exchanged with the use 
%of syntax. 
of compositionality, a rudimental form of syntax.
In this situation, the concept is represented by 
a  couple of symbols $\alpha_x+\beta_y$, each one sorted from 
a different inventory $K_x$ and $K_y$.
It follows that each agent is characterized by two different 
inventories $K_x$ and $K_y$.
At each time step, the following microscopic 
rules control the communication:

1a) The speaker retrieves a word ($\alpha_x$)
from its inventory $K_x$; if its inventory is empty, the speaker invents a new word.

1b) The speaker retrieves a word ($\beta_y$)
from its inventory $K_y$; if its inventory is empty, the speaker invents a new word.

2) The speaker transmits the selected pair of words to the hearer.

3a) If the hearer's inventories $K_x$ and $K_y$  contain the pair of words 
$(\alpha_x,\beta_y)$, the communication is a success. 
The two agents update their inventories so as to keep, in each one, 
only the correspondent words involved in the 
interaction ($\alpha_x$ in $K_x$ and $\beta_y$ in $K_y$).
 
3b) Otherwise the communication is a failure. 
The hearer adds the words he does not know 
(one or two) to the corresponding inventory/ies;
the speaker does nothing.\\

Here we hypothesize that the learning of segmented elements
of the utterance is possible even if the communication is a failure.
This idea is supported by the fact that it is not necessary to have any 
positive feedback to identify the components of a speech.
In fact, some popular experiments shown how very young infants 
can achieve  the task of word 
segmentation of an utterance with only minimal exposure, 
just by exploiting the transitional probabilities between syllables \cite{safran}.
Even so, the ability to use exclusively statistical information coming from a passive 
exposure to process a given language  stream seems to 
be confined to the individuation of the segments of a stream, but not to acquiring
the generalization correspondent to a syntax structure \cite{pena}. 
For this reason, we make the hypothesis that the fixation of the structured 
element $\alpha_x+\beta_y$ is possible only if there
is a communication success. Which means, on the basis of a 
exposure to a positive feedback.

\subsection{Many concepts games}

In this situation agents develop communications which 
can exchange $C$ different concepts.
If no syntax structures are present, each concept is represented by one symbol.
If syntax is introduced, a combination of two symbols, $\alpha_x+\beta_y$,
each one picked up from a different inventory, represents each concept.
As in the basic model, each symbol is represented by one of 
$32$ different possible words.

It follows that for a $C$ concepts Game,  if there is no syntax, 
each agent is represented by $K_z,z=1,2,..,C$ inventories, 
each one containing no more than $32$ words. 
These words are exchanged in the same way as in
a single concept Naming Game without syntax.

\begin{figure}
\begin{center}
\includegraphics[width=0.5\textwidth, angle=0]{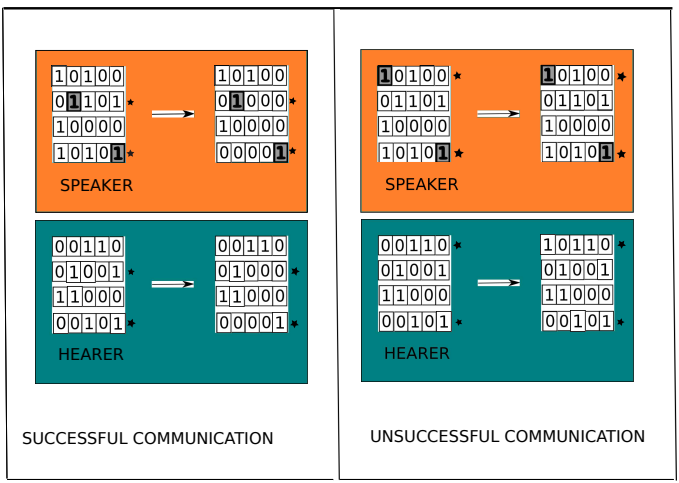}
\end{center}
\caption{\small (Color online) 
An example of the structure and evolution of the inventories 
for the model with syntax when four concepts ($S=2$) are exchanged. 
In this particular example the number of inventories ($2S$)
is equal to the number of concepts. 
%On the left we present a successful communication.} 
We can think that the first two inventories represent nouns and
the third and fourth stand for verbs.
The starred inventories are the ones corresponding to the concept sorted out
in a specific communication event. 
The shadowed elements are the ones transmitted by the speaker. 
}
\label{fig_draw}
\end{figure}

In the case of a syntactic communication, 
as explained in the introduction, 
the concepts $C_{ij}$ are represented by the elements,
generated by the combination of two symbols, 
of a matrix $C_{ij}=S\times S$.
It follows that a $C$ concepts Game with syntax is obtained by
introducing agents
represented by $K_z$, $z=2,..,2S$ inventories, 
each one containing no more than $32$ words.
In Table~\ref{tab_Con} we give an example of a 
Game with nine concepts. 
At each time step a concept is chosen 
determining the two inventories that represent it 
(for example $K_a$ and $K_e$ from which we represent the concept with
the couple $\alpha_a+\beta_e$).
The dynamics of each single communication are the same as 
the one concept game with syntax
(an example of a four concepts game is represented in Figure~\ref{fig_draw}).

\section{Results and discussion}

\subsection{One concept model}

We describe the time evolution of our system looking at some  
usual global quantities \cite{baronca,edo1,edo2}: the total number of words ($N_{tot}$) 
present in the population %, the number of different words ($N_{dif}$) 
and the success rate ($R_S$), which measures the average 
rate of success of communications. This is obtained by evaluating
the frequency of successful communications in a given time interval.

The basic model is a simplified version of the original 
Naming Game where the number of different words introduced in
the system is limited and sorted among a cache of $32$.
As a result, a homologous behavior occurs with some limited differences 
(see also \cite{edo1} for similar results). 
In our version of the model the maximum number of words for each agent is constant and it does not scale with the square root of the population. 
This important fact is responsible for %triggering the changes 
remodeling the temporal scaling behavior.
Anyway, it is important to point out that, as the mean number of words for each player is always well below $32$, the model maintains all the basic features of the original one and the possibility of the invention of new words is not affected.

\begin{figure}
\begin{center}
\includegraphics[width=0.5\textwidth, angle=0]{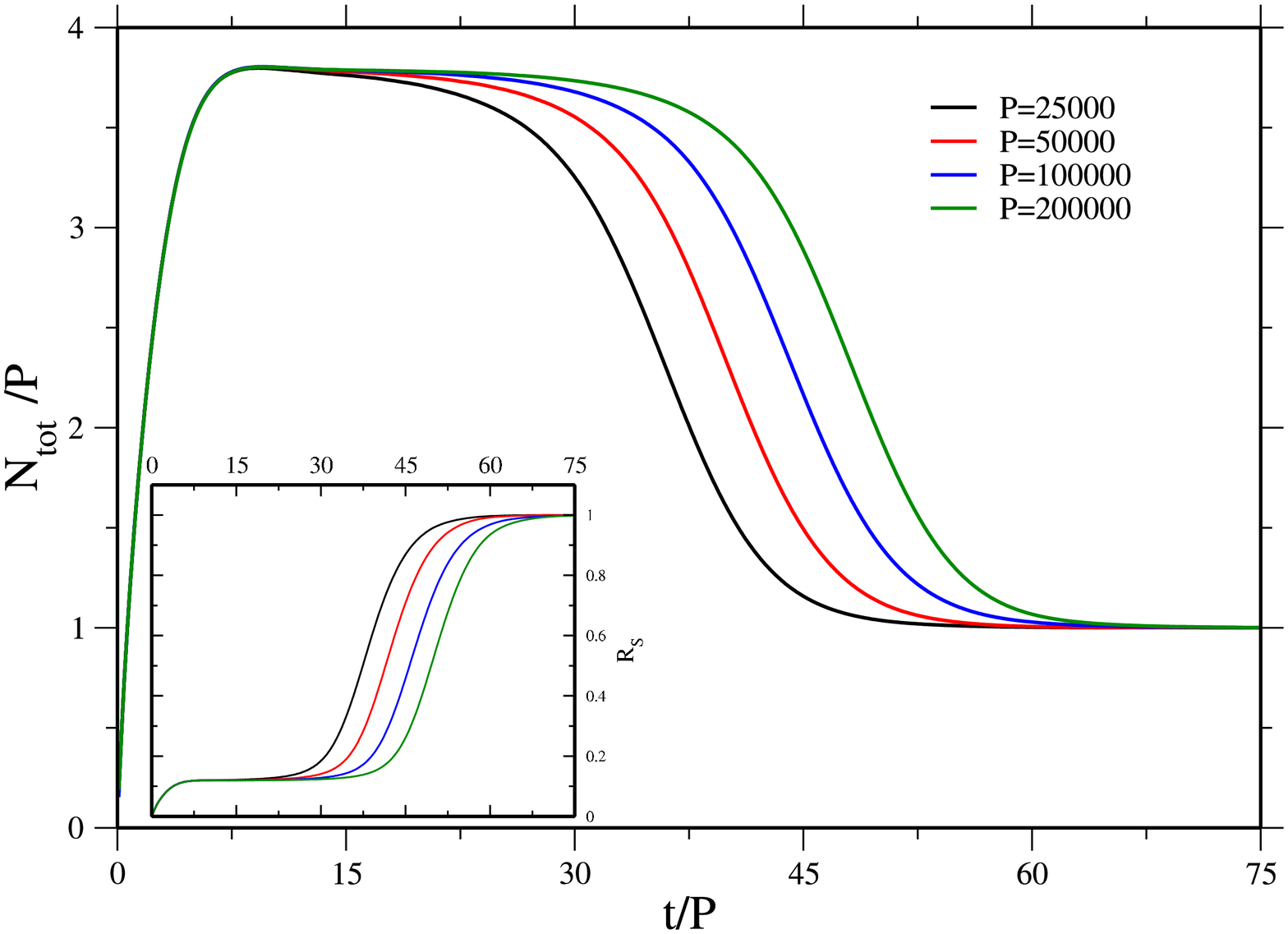}
\includegraphics[width=0.5\textwidth, angle=0]{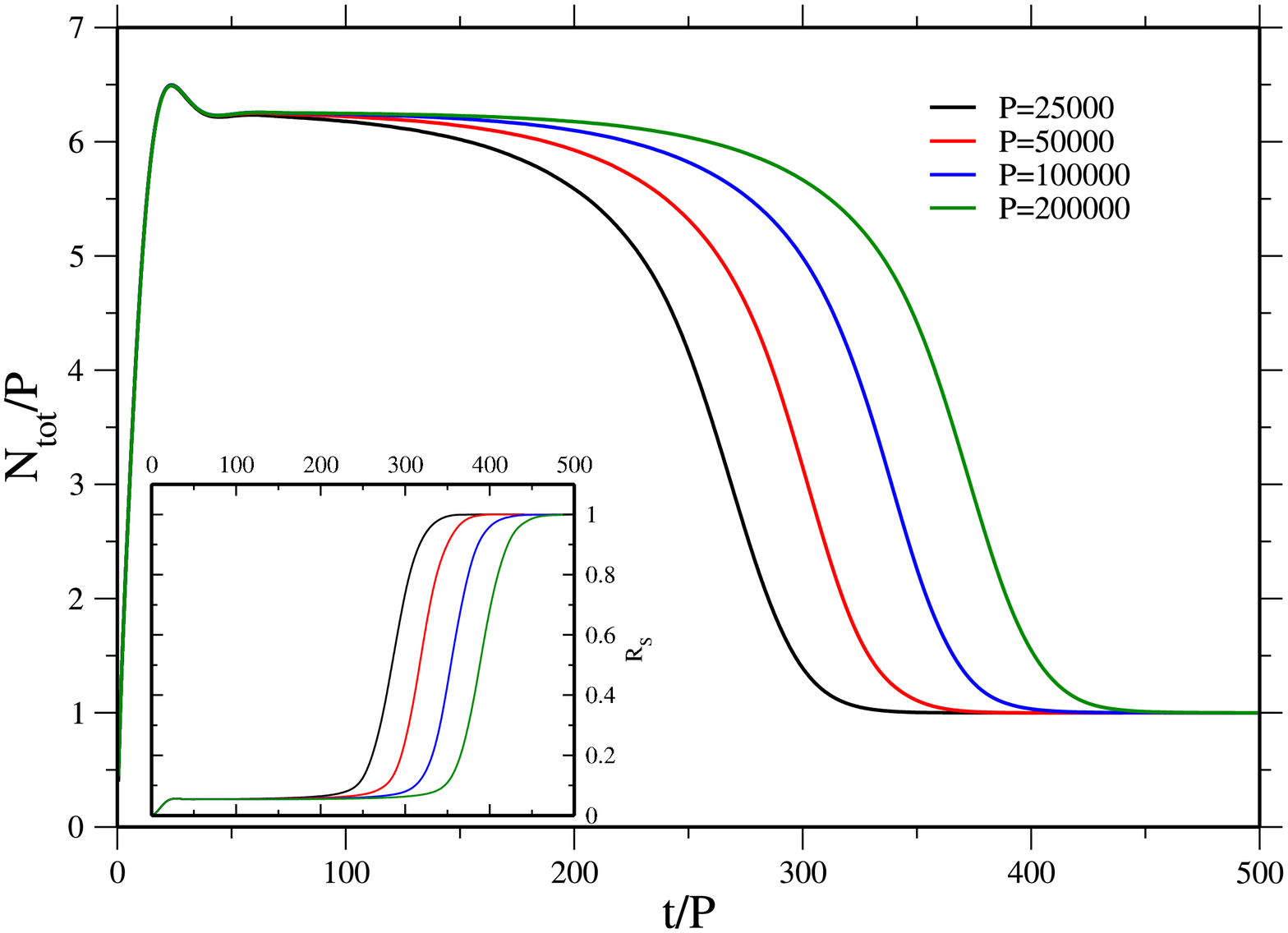}
\end{center}
\caption{\small(Color online)  One concept model.
Top: Model without syntax. We present the temporal evolution for the total number of words 
divided by the total population. 
The inset shows the success rate $R_S(t)$. 
Bottom: The same data for the syntactic model. 
All data are averaged over 1000 simulations. 
%S-shaped curves in language change -old linguistic variants replaced by news ones s-shaped curves in plots of frequency against time (C. Yang Lang. Var. Change, 12 231 2001, L. Pearl and A. Weinberg, Lang. Learn. Dev. 3, 43 2007)
% (documented in numerous studies of language change (A. Kroch, Lang. Var. Change 1, 199 ,1989)
}
\label{fig_pheno}
\end{figure}

All agents start with an empty inventory
and an initial transient exists 
which corresponds to the rise of $N_{tot}(t)$.% and $N_{dif}$.
This stage finishes when this quantity attains a maximum value
($Max[N_{tot}]\approx 3.8 P$) which is maintained along a plateau.   
When the redundancy of words reaches a sufficiently high level, 
the number of successful plays increases. 
The curve $N_{tot}(t)$ %and $N_{dif}(t)$ 
begins a decay 
towards the consensus state, corresponding to 
one common word for all the players, reached at time $T$.
In Figure~\ref{fig_pheno} we report the temporal evolution for $N_{tot}(t)$
and $R_S(t)$.
% $N_{dif}(t)$ and $S(t)$. \\

As shown in \cite{baronca}, it is possible to estimate the 
maximum number of total words using some simple analytical considerations.
If we represent  the mean total number of words for an
agent, at time step $t$, with $n(t)$, and  the mean total number 
of different words with $D(t)$ we obtain:

\begin{eqnarray}
\label{eq_1}
n(t+1)-n(t)&=& \frac{1}{n(t)}\Big(1-\frac{n(t)}{D(t)}\Big)\frac{1}{2}%+\\ \nonumber
-\frac{1}{n(t)}\frac{n(t)}{D(t)}(n(t)-1)\nonumber \\
\end{eqnarray}
We are considering that the probability for the speaker to communicate 
a specific word is $\frac{1}{n(t)}$ and 
the probability for the hearer to own that word is $\frac{n(t)}{D(t)}$. 
It follows that the first term represents the gain term for a 
failed communication (which increases $n(t)$ by $1/2$), 
and the second term represents the loss term (which decreases $n(t)$ by $n(t)-1$). 
We can use this equation for describing the $P$ dependence.
If we assume that at the plateau, where we can consider  $n(t+1)-n(t)=0$, 
$n(t)$ scales as $\alpha P^{\beta}$, and that 
% for large $P$, 
$D\approx 32$, we can write:

\begin{eqnarray}
\label{eq_2}
\frac{1}{2\alpha P^{\beta}}\Big(1-\frac{\alpha P^{\beta}}{32}\Big) = \frac{1}{32}(\alpha P^{\beta}-1)
\end{eqnarray}

This equation reduces to $\frac{1}{P^{\beta}} \propto P^{\beta}$, 
which forces $\beta=0$. 
This fact implies that  the number of total words for each 
player is not dependent on $P$.
It follows that $Max[N_{tot}]\propto P $, as can be seen in Figure~\ref{fig_pheno}.
Equation \ref{eq_2} can also be used to 
evaluate the exact numerical value of the plateau. 
It is sufficient to consider $\alpha P^\beta$ as a constant 
and the corresponding value is $4.25$.
If we take into consideration that our equations are a mean field 
approximation which does not account for the 
correlations builded up between the individuals' inventories, the 
value is comparable with the result $3.8$ obtained by the simulations. \\

\begin{figure}
\begin{center}
\includegraphics[width=0.5\textwidth, angle=0]{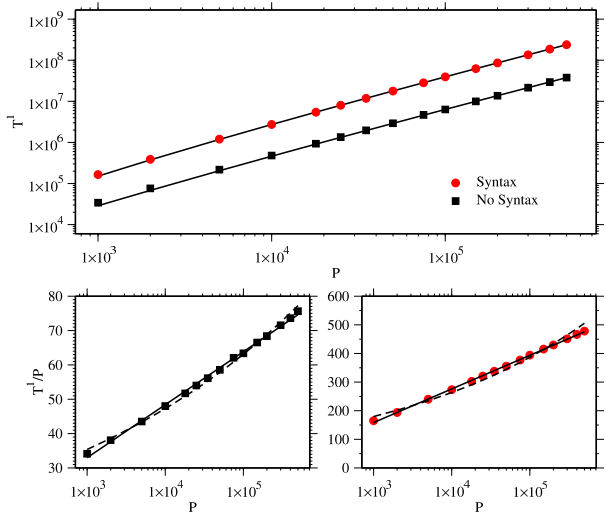}
\end{center}
\caption{\small 
We present the convergence time as a function of the population size for the one concept model.
Top: 
The model with no syntax is well fitted by the depicted relation: 
$T^{1}_{ns} \approx -22.1 P+7.6 P\ln{P}$.
The model with syntax is well fitted by the relation:
$T^{1}_{s} \approx -213.2 P+52.7 P\ln{P}$. 
Bottom:  Rescaled convergence times for the model without syntax (on the left) and with syntax (on the right).
The rescaled times are well fitted by the function $a_1+a_2 \ln{P}$ (continuous lines).
A power law fit of these data ($b_1 P^{b_2}$), represented by the dashed lines, turns out
to be moderately less accurate than the previous one, which can be derived from some theoretical considerations.
}
\label{fig_scalingP}
\end{figure}

We explored the behavior of the convergence time $T^{1}_{ns}$ 
(the index $1$ stays for a one concept play). 
As stated before, this is the time at which the system reaches 
the consensus state, corresponding to 
one shared word for all the players. 
We studied the dependence of $T^{1}_{ns}$ on $P$ averaging over 
different simulations obtaining, throughout a regression, the following 
dependence: $T^{1}_{ns}(P) \approx -22.1 P+7.6 P\ln{P}$ 
(see Figure~\ref{fig_scalingP}).
This result corresponds to the average convergence time over 
different simulations, which is obviously different from 
the convergence time of the mean simulation (the one presented in
Figure~\ref{fig_pheno}).
These analyses, as well as the following ones, are consistent 
%valid/make sense only
for sufficiently large populations.

We can support these numerical results 
with some analytical considerations analogous to the ones
presented in \cite{baronca2}. 
During the time evolution we can distinguish two periods. 
A first interval, between $t=0$ and the time when the system reaches the maximum number of words ($t_{max}$), which clearly scales linearly with the population size (see Figure~\ref{fig_pheno}).
A second interval, between $t_{max}$ and $T_{ns}^1$, which 
is governed by the following dynamics.
To reach convergence, the mean number of
words for each individual, which does not depend on $P$, has to decrease to $1$.
As at each play the loss term does not depend on $P$, 
from the definition of the dynamics of our model
a necessary condition for convergence is that each agent 
must win at least once.
For this reason, near convergence, the number of agents 
which did not have a successful interaction ($P^*$) should be finite.
We can estimate this number. In fact $P^*(t)=P(1-R_S(t)/P)^t$, where $1/P$ 
is the probability of selecting an agent and $R_S(t)$ corresponds to the probability of a success . 
As can be appreciate in the inset of Figure~\ref{fig_pheno}, 
$R_S(t)$ does not depend on $P$ and it is practically constant for a long time after $t_{max}$.
It follows that $(1-1/P)^{t_{diff}}\propto1/P$, where $t_{diff}=T^{1}_{ns}-t_{max}$.
For large $P$ we obtain: $t_{diff} \propto P \log(P)$.
This condition turns out to be sufficient when confronted with the numerical data.
In fact, $T^{1}_{ns}$ results to be very well fitted by a function of the type 
$c_1P+c_2Pln{P}$. \\

We can unfold a similar analysis for the syntactic 
model with one concept.
In this case we are implementing a Game where each agent 
is represented by
two inventories which follows the rules presented in the previous section.
From the results of our simulations we can observe that the crucial features that
determine the scaling properties of the convergence times are also maintained in this
scenario. In fact, as can be seen in Figure~\ref{fig_pheno}, 
$Max[N_{tot}]\propto P$, $t_{max}$ scales linearly with $P$ and $R_S(t)$ 
continues independent of P.
As the same arguments produced for 
estimating $T^1_{ns}$ continue to be valid,
we can expect the $T^{1}_{s}$ dependence on 
$P$ to have the same functional form of $T^{1}_{ns}$.
Fitting our numerical data with such function (see Figure~\ref{fig_scalingP})
we obtain the following scaling relation for the 
convergence time for a syntactic model with one concept:
$T^{1}_{s}(P) \approx -213.2 P+52.7 P \ln{P}$.

\begin{figure}
\begin{center}
\includegraphics[width=0.5\textwidth, angle=0]{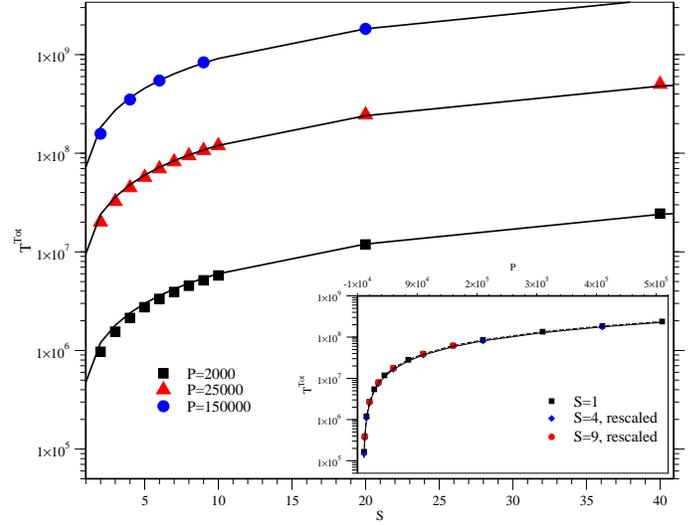}
\end{center}
\caption{\small (Color online)
The convergence time as a function of $S$ for different populations.
The linear regressions are very
well approximated by the general expression $T_s^{Tot}\approx1.5 S T_s^1(P)$ (the continuous lines).
The inset shows the convergence time as a function of $P$ for different $S$ values. The convergence times for $S=4,9$ are rescaled by the factor $1.5 S$. The continuous lines are the rescaled curves $c_1 P + c_2 P \ln{P}$ obtained from the regressions. The rescaling transformation produces a good collapse of  data and curves.
Data are averaged over up to 1000 simulations.
}
\label{fig_scalingS}
\end{figure}

\subsection{Many concepts model}

First, it is useful to introduce a new quantity, the total 
convergence time $T^{Tot}$.  
This is the time when the system reaches 
the consensus state, corresponding to 
one shared expression for each concept and for all the players.
Equivalently, it is the time when every communication event, 
relative to any one of the possible concepts, is a success.
 
Given an experiment set up so that $C$ concepts are exchanged, 
if there is no syntax, the behavior described for the basic Game with no syntax
is reproduced for each one of the concepts, and so it is easily generalized.
In fact, the dynamics of each concept is obviously 
independent of those of the other concepts.
This fact implies that if there is no syntax 
$T_{ns}^{Tot}=CT^{1}_{ns}(P)=S^2T^{1}_{ns}(P)$.

This is not the case if we introduce syntax.
In this case, as we stated before,  we consider the situation
in which the matrix $C_{ij}=S\times S$ represents all the exchanged concepts.  
We explored numerically the behavior of the total convergence time as a function of the number of concepts.
We assume as a null hypothesis that $T_s^{Tot}$  depends linearly on 
the dimension of the matrix $C_{ij}$.
As presented in Figure~\ref{fig_scalingS} the analysis of the data from simulations
suggests that this scaling relation is satisfied.
Moreover, for $S\ge2$, we can express the regression coefficients for different population sizes with a simple expression which uses $T_s^1(P)$: %equation \ref{eq_3}:
$T_s^{Tot}\approx 1.5 S T_s^1(P)$.
This expression is derived from a numerical analysis of the data obtained from simulations.
Some results supporting this scaling are reported in Figure~\ref{fig_scalingS}.

Starting from the scaling relation for the convergence time in dependence on $P$
and $S$ we can determine the different behavior generated by the introduction
of syntax.
Quantifying this convergence time allows us to determine the strategy that enables 
a more effective communication. In fact, reaching 
consensus at a collective level corresponds to an efficient 
communication at an individual level.
Depending on the number of concepts exchanged and on the 
population size we can determine if the total convergence time 
for a model with syntax is shorter than that for a model without syntax.
This situation is attained if   $1.5 S T_s^1(P) <S^2 T^{1}_{ns}(P) $.
Using this estimation we are able to determine 
a critical value of $S$ for which the emergence of 
syntax is viable: $S>1.5 T_s^1(P)/T^{1}_{ns}(P)$.

\begin{figure}
\begin{center}
\includegraphics[width=0.5\textwidth, angle=0]{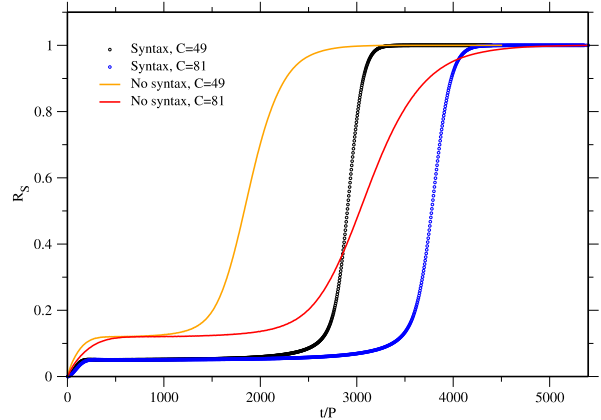}
\end{center}
\caption{\small(Color online)
Temporal evolution of the success rate for a different 
number of exchanged concepts (C) for a model without syntax or with syntax. 
For  $C\ge81$,  $T_{s}^{Tot}<C T^{1}_{ns}$, 
which means a faster convergence for population using syntax. $P=25000$}
\label{fig_emergence}
\end{figure}

From this relation it follows that, if the number of exchanged concepts 
is sufficiently large in relation to the population size,
the syntactic model is able to generate a faster convergence towards consensus.
The dependence on the population size is very weak. 
For a population of 2000 individuals 
$S=8$ is sufficient for the conventionalization of syntax,
and for a population 100 times larger it is sufficient to select $S=10$. 
So, from an empirical point of view, for typical populations, 
the relevant factor is simply the number of exchanged concepts, 
an interesting fact that enhances the possibility of syntax to emerge as 
an auto-organized process.
The results obtained using our approximation were confirmed by 
different numerical simulations.
In Figure~\ref{fig_emergence} we present an example  for $P=25000$.
In this case the 
matrix dimension should be bigger than $8$ and effectively, from our simulation, if we exchange $81$ concepts ($S=9$)
the syntactic model clearly performs better than the non syntactic one.
In others words, the introduction 
of syntax generates a  social communicative
advantage when language must cope with a lot of concepts and 
when it is employed in smaller communities.
In this context, the transition from non-syntactic to syntactic communication 
is socially favored. 
 
\section{Conclusions}

We formulate a simple framework to explore 
the possibility of the emergence of an elementary syntactic structure
fixed by the social dynamics defined by communication.
We start with a version of the Naming Game model
generalized for exchanging many different concepts.
A simple syntactic structure is introduced in the form of a binary combination process
and the algorithm for fixing this structure is inspired by some known results
relative to individual learnability of linguistic structures.
In this way we can analyze the transition between syntactic and non syntactic 
communication on the basis of the social communicative potential of a linguistic structure,
and not on the basis of the
individual fitness or the velocity of individual learning.

From the analysis of this model, we can show that, 
under certain conditions, syntactic communication can reach consensus 
more efficiently than non syntactic communication, even if the the task of fixing a syntactic structure is more difficult.
We are able to  evidence some critical values 
for the number of exchanged concepts
in dependence on the population size for which the emergence 
of syntax is viable. 

\section*{Acknowledgments} 

I am grateful to Marcio Argollo de Menezes 
and David Eduardo Zambrano Ram\'irez
for fruitful discussions.
I thank the anonymous referees for constructive and helpful comments.

\end{multicols}


\section*{References}

\begin{thebibliography}{99}

\bibitem{chomsky1}        
N. Chomsky, {\it Rules and representations}, Blackwell, Oxford (1980).
\bibitem{nowak}
N.L. Komarova and M.A. Nowak, Proc. R. Soc. Lond. B {\bf 268}, 1189 (2001); M. A. Nowak, N. L. Komarova and P. Niyogi, Nature {\bf 417} , 611 (2002).

\bibitem{chomsky05}
N. Chomsky, Ling. Inq. {\bf 36}, 1 (2005).        
        
\bibitem{chater1}        
N. Chater, F. Reali and M.H. Christiansen, PNAS, {\bf 106}, 1015 (2009).	

\bibitem{hsu}
A.S. Hsu, N.Chater and P. Vit\'anyi, Cognition, {\bf 120}, 380 (2011).

\bibitem{pearl}
L. Pearl and A. Weinberg, Lang. Learn. Dev., {\bf 3}, 43 (2007).

\bibitem{christiansen}
M.H. Christiansen and N. Chater, Behav. Brain Sci. {\bf 31}, 489 (2008).

\bibitem{reali}
F. Reali and T.L. Griffiths, Proc. R. Soc. B {\bf 277}, 429 (2010).         

\bibitem{steels1}
L. Steels, {\it The Talking Heads Experiment}, Vol.1 Laboratorium, Antwerpen (1999). 

\bibitem{steels2}
L. Steels, Artificial Life {\bf 2}, 319 (1995).
L. Steels, Trends. Cogn. Sci. {\bf 7}, 308 (2003).

\bibitem{baronca} A. Baronchelli, M. Felici, E. Caglioti, V. Loreto and L. Steels, 
J. Stat. Mech.,  P06014 (2006).

\bibitem {gong} A. Baronchelli, T. Gong, A. Puglisi, and V. Loreto, PNAS  {\bf 107}, 2403 (2010). 

\bibitem{puglisi} A. Puglisi, A. Baronchelli, and V. Loreto, PNAS  {\bf 105}, 7936 (2008).

\bibitem{bicke1}
D. Bickerton and E. Szathmáry, {\it Biological Foundations and Origin of Syntax}, MIT Press, Cambridge MA (2009).

\bibitem{chomsky3}
N. Chomsky, {\it The Minimalist Program}, MIT Press, Cambridge MA (1995).

\bibitem{progovac}
L. Progovac and J.L. Locke, Biolinguistics {\bf 3}, 337 (2009).

\bibitem{bickerton}
D. Bickerton, {\it  Bastard Tongues},  Hill \& Wang, New York  (2008).

\bibitem{nowak2}
M.A. Nowak, J.B. Plotkin and V.A.A. Jansen, Nature {\bf 404}, 495 (2000).

\bibitem{safran}
J.R. Safran et al. Science, {\bf 274}, 1926 (1996).

\bibitem{pena}
M.Pe\~na, L.L. Bonatti, M. Nespor and J. Mehler, Science, {\bf 298}, 604 (2002).

\bibitem{edo1} 
E. Brigatti and I. Roditi, New Journal of Physics, {\bf 11}, 023018 (2009).

\bibitem{edo2} 
E. Brigatti, Phys. Rev. E, {\bf 78}, 046108 (2008).

\bibitem{baronca2} 
A. Baronchelli, V. Loreto and L. Steels, Int. J. of Mod. Phys. C, {\bf 19}, 785 (2008). 


\end{thebibliography}
\end{document}